# Fermi level tuning and band alignment in Mn doped InAs/GaSb

L. Riney,[1] J. Bermejo-Ortiz,[2] G. Krizman,[3] S.-K. Bac,[1] J. Wang,[1] M. Zhukovskyi,[4] T. Orlova,[4] L.A. de Vaulchier,[2] Y. Guldner,[2] R. Winkler,[5] J.K. Furdyna,[1] X. Liu,[1] B.A. Assaf[1]

[1] Department of Physics, University of Notre Dame, Notre Dame IN, 46556

[2] Laboratoire de Physique de l'Ecole Normale Supérieure, ENS, Université PSL, CNRS, Sorbonne Université, 24 rue Lhomond 75005 Paris, France

[3] Institut für Halbleiter- und Festkörperphysik, Johannes Kepler Universität, Linz, Altenberger Strasse, 69, 4040 Linz, Austria

[4] Notre Dame Integrated Imaging Facility, University of Notre Dame, Notre Dame IN, 46556

[5] Department of Physics, Northern Illinois University, DeKalb IL

**Abstract.** InAs/GaSb hosts a broken gap band alignment that has been shown to generate helical topological edge states. Upon the introduction of Mn into the structure, it has been predicted to host a quantized anomalous Hall effect. Here, we show that dilute Mn doping on InAs in InAs/GaSb, allows a tuning of the Fermi level, the introduction of paramagnetism, but also has a non-trivial impact on the band alignment of the system. The measurement of Shubnikov-de-Haas oscillations, cyclotron resonance, and a non-linear Hall effect in Mn-doped samples indicate the coexistence of a high mobility two-dimensional electron gas and a hole gas. Conversely, in undoped InAs/GaSb, pure-n-type transport is observed. We hypothesize that Mn acceptor levels can pin the Fermi energy near the valence band edge of InAs, far from the interface, which introduces a strong band bending to preserve the band offset at the InAs/GaSb interface. The realization of the QAHE in this structure will thus require a careful control of the band alignment to preserve topological insulating character.

I. Introduction

The InAs/GaSb broken-gap III-V heterojunctions [1] can host a two-dimensional quantum spin Hall state resulting from a band-inverted electronic structure [2] [3] [4] [5]. At the interface between the two materials, the InAs conduction subband edge lies below the GaSb valence subband edge for a given thickness range. [6] The two energy bands can hybridize and open a gap. [7] [8] [9] Thus, this interface can host two-dimensional electron states HH1 and E1 that have a topological band-inverted gap. [2] Various experiments have reported evidence of this band inversion or of edge states in this structure. [10] [11] [12] [13] Magnetic exchange induced onto the energy states of this system can have exciting consequences that have not been investigated. Specifically, it was predicted to yield a quantized anomalous Hall effect (QAHE), by inducing a Zeeman splitting that restores trivial band order for one spin but maintains the inversion for the other. [14] This state has yet to be observed but can only be reached if the Fermi level lies between HH1 and E1 and the system has a finite magnetization. In a sample that satisfies these conditions and maintains a bulk insulating character, it is expected that chiral edge states appear without the need for Landau quantization. [15]

A realistic mapping of the topological phase diagram of InAs/GaSb with magnetism introduced by the presence of Mn, also requires one to preserve Landau quantization. This is needed primarily to allow a reliable determination of the strength of magnetic exchange coupling compared to the inverted gap between HH1 and E1 using various Landau level spectroscopies. [16] [17] It is thus important to properly quantify the extent to which Mn alters the diffusive and quantum properties of the interfacial 2DEG of InAs/GaSb.

While $In_{1-x}Mn_xAs$ compounds have been extensively studied in the past, those studies were limited to quantum wells, epilayers and heterojunctions with high Mn content. Digital doping of InAs with Mn was also studied. [18] The introduction of a high concentration of Mn (>1%) into InAs/GaSb was studied to enhance magnetism and the anomalous Hall effect in the past; however, the impact of Mn doping on quantum transport and on the band alignment in InAs/GaSb was never considered. [19] [18] Given that Mn plays a doubly important role of introducing magnetism and acceptor levels in III-V materials, it is expected to have a non-trivial impact on the system. In bulk InAs, Mn impurity levels are known to occur near the valence band edge, leading to p-type conduction, with low mobility. [19] [20] However, their behavior was yet not investigated in InAs/GaSb when introduced at dilute levels.

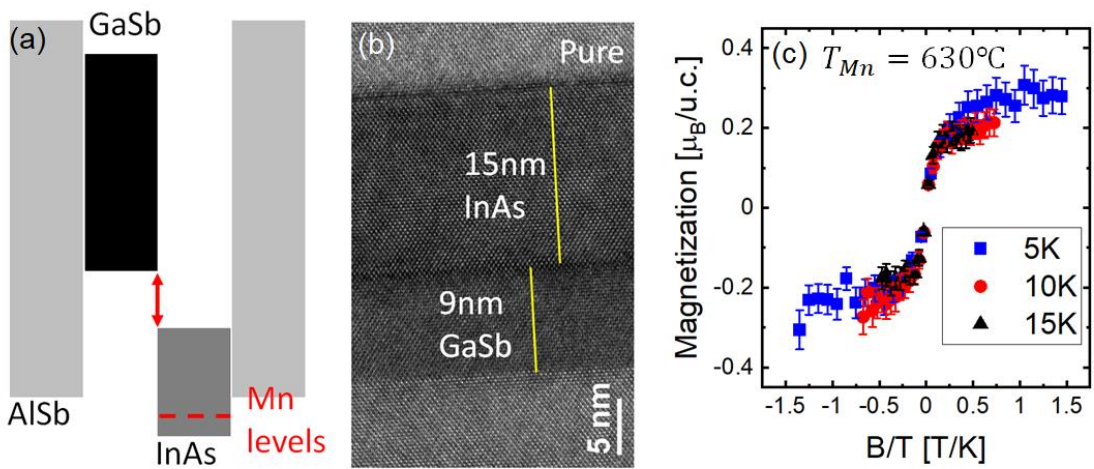

**FIG 1**. (a) Band alignment diagram of InAs/GaSb bilayers grown between AlSb barriers. The position of the Mn levels is hypothetical. (b) Transmission electron microscopy images of a pure bilayer showing the thickness of each layer. (c) Magnetization versus B/T at 5K, 10K and 15K with magnetic field applied along the growth axis for sample 630. The error bars are based on the statistical standard deviation at saturation.

Here, we study the electrical and magnetic properties of paramagnetic $In_{1-x}Mn_xAs$/GaSb at low x (<0.1%) grown by molecular beam epitaxy. We consider the impact of Mn on the Fermi energy and band alignment of the system. Quantum oscillations are measured up to magnetic fields reaching 16T. We observe persistent quantum oscillations from the interfacial two-dimensional electron gas (2DEG) of InAs/GaSb coexisting with p-type conduction and paramagnetism in the presence of Mn. The frequency of the oscillations and the total charge density of the 2DEG decrease with increasing Mn content. However, a hole gas starts to emerge as Mn content or temperature are increased, without altering the qualitative behavior of the interfacial 2DEG. We thus conclude that Mn introduces a significant band bending into the system which indicates that a band structure calculation that is self-consistent with the electrostatics of the system is required to fully understand the properties of $In_{1-x}Mn_xAs$/GaSb. The realization of the QAHE in this system could rest on the proper understanding and control of the band alignment.

| Mn Cell [°C] | Sheet resistance [Ω] | Low field Hall slope [Ω/T] | Hall Carrier density [cm$^{-2}$] | SdH Carrier density [cm$^{-2}$] | Mn content x |
|---|---|---|---|---|---|
| 0 | 267 | -93 | - 6.7E12 | - 6.7E12 | 0 |
| 600 | 48 | -88 | - 7.1E12 | - 6.5E12 | <10$^{-4}$ |
| 615 | 120 | -73 | Two carrier types | - 4.9E12 | <10$^{-4}$ |
| 630 | 196 | -63 | Two carrier types | - 3.0E12 | *0.0013±0.0002 |

Table I. Mn cell temperature also used as sample ID, sheet resistance, Hall effect slope R$_H$, the carrier density extracted from Hall and Shubnikov-de-Haas (SdH) oscillations, and the Mn content extracted from magnetometry where possible. For samples 600 and 615, the order of magnitude of x is estimated from the carrier density. *For 630, it is measured using SQUID magnetometry on a multiquantum well of 10 periods of AlSb/InAs/GaSb grown under the same conditions as the double wells studied here.

## II. Results

**Growth and characterization**

In$_{1-x}$Mn$_x$As/GaSb wells are grown by molecular beam epitaxy on GaAs(100). After substrates are prepared, a series of strain-relieving buffer layers is grown as follows: 250nm AlSb, 500nm(Al,Ga)Sb, and 22nm AlSb as a barrier. This is followed by the GaSb well, the InAs well, a 22nm AlSb top barrier, 95nm(Al,Ga)Sb, and a 2-3nm GaSb capping layer to prevent the oxidation of Al containing layers. Mn is incorporated by co-deposition during the InAs growth. The structure is sketched in Fig. 1(a) in the vicinity of the wells. Throughout this manuscript, the nominal Mn content is denoted by the Mn cell temperature. It is also used for sample identification. The sample properties are summarized in Table 1. Fig. 1(b,c) shows a high-resolution transmission electron microscopy (TEM) image of the InAs/GaSb interface of the pure sample. The abrupt nature of the interface is evident in the TEM images. Those images also yield a precise thickness of 15nm for InAs and 9nm for GaSb. In table I, we list 4 samples that we study. Each sample will be referred to by the Mn cell temperature that was utilized during the growth.

SQUID magnetometry measurements were carried out on a multiquantum well with 10 periods grown under the same conditions as sample 630. The 10 periods are required to enhance the paramagnetic signal. Fig. 1(c) shows the magnetization plotted versus B/T for this sample, at 5K, 10K and 15K. A clear s-shaped curve is recovered characteristic of the paramagnetism. The magnetization saturates close to $(0.26 \pm 0.04)\mu_B/u.c.$ This yields the Mn content shown in table I for sample 630. For a lower Mn cell temperature, the paramagnetism due to Mn atoms remained below our detection capabilities. However, as will be seen later, the presence of Mn is confirmed by the changing charge density of the system.

Magnetotransport measurements are carried out at 1.5K in a standard Oxford Instruments cryostat equipped with a superconducting magnet that can reach 16T. Rectangular samples are cleaved from large wafers and bonded using In solder or silver paint to obtain 5 contacts. Magnetooptical infrared spectroscopy measurements are carried out at 4.2K up to 15T using a cryostat system coupled with a Bruker Vertex 80v spectrometer configured for the far-infrared range. A bolometer cooled down to 4.2K in the same bath as the sample is used to detect optical transmission T through the sample. The relative transmission T(B)/T(B=0) is reported as a function of field.

**Quantum oscillations at various Mn contents**

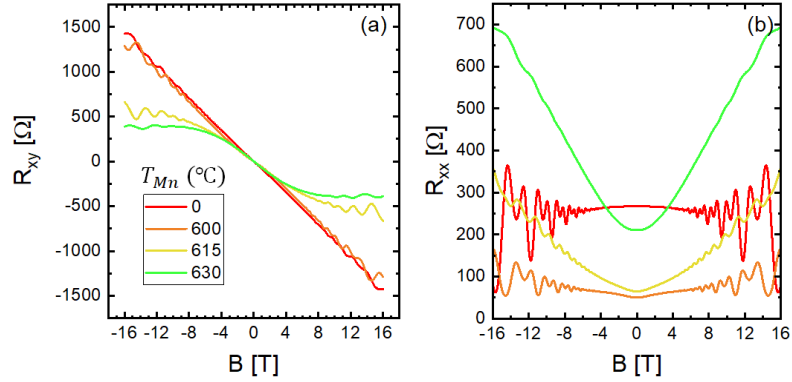

**FIG 2**. (a) Hall effect and (b) Magnetoresistance measured in all 4 samples at 1.5K for various Mn cell temperatures.

Figure 2 shows the variation of the Hall effect measured in the 4 samples grown at varying Mn cell temperatures, including an undoped sample. All exhibit an n-type Hall effect at low field and host strong quantum oscillations from Landau quantization. In Fig. 2(b), we plot the magnetoresistance (MR) measured at 1.5K for the 4 samples. These initial measurements allow us to extract the diffusive transport properties of the samples that exhibit a quasi-linear Hall effect. They are listed in Table 1. A reduction of the Hall effect slope at low magnetic field correlates with an increasing Mn cell temperature at which the samples were grown. The amplitude of the quantum oscillations shown in Fig. 2(b) also drops with increasing Mn content.

Fig. 3(a) shows the oscillating resistance after a background is removed. A Fourier transform of the curves is performed and is shown in Fig. 3(b). It allows us to extract the oscillation frequencies as well as the n-type carrier density of the samples. We initially focus on the undoped sample, for which we obtain two frequencies that yield a $n_1 = 4.2 \times 10^{12} cm^{-2}$ and $n_2 = 2.3 \times 10^{12} cm^{-2}$. Their sum agrees with what we extract from the Hall effect slope shown in Fig. 2, $n_H = 6.7 \times 10^{12} cm^{-2}$. We can further understand the origin of this by carrying out self-consistent calculations in the Hartree approximation using an envelope function model based on the 8×8 Kane Hamiltonian. [21] [22] [23] We model the InAs/GaSb system, by taking into account the established band offset between the two materials $E_{V,GaSb} - E_{C,InAs} = 0.15eV$. [24] The elevated position of the Fermi level above the conduction band edges is likely due to bulk impurities in the InAs well, as confirmed by Hall measurements carried out on a control epilayer. Thus, we assume that the quasi-2D charge density in the system is due to a homogenous distribution of donors in the InAs well that implies overall charge neutrality.

The model yields the weakly spin-split energy band diagram shown Fig. 3(c), where we have marked the position of the Fermi level that agrees with the experimentally determined charge densities for the undoped sample. Indeed, the Fermi level crosses two electron-like bands, yielding Fermi wavevectors that reproduce the densities measured from quantum oscillations (calculated subband densities $n_1 = 4.3 \times 10^{12} cm^{-2}$ and $n_2 = 2.5 \times 10^{12} cm^{-2}$). A theoretical model that assumes flat-band conditions thus reproduces the findings of the experiments at 1.5K, in the absence of Mn.

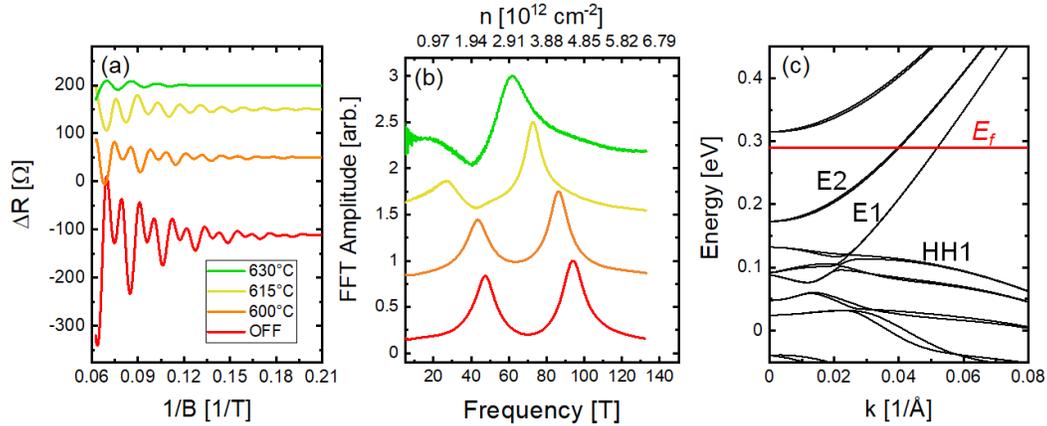

**FIG 3.** (a) Shubnikov-de-Haas oscillations measured in 4 samples labeled by the Mn cell temperature during their growth, after a background is removed. (b) FFT power spectra of the oscillations shown in (a). The spectra are normalized by the largest peak amplitude. The curves in both (a) and (b) are shifted for clarity. (c) Energy band diagram of the InAs/GaSb system calculated assuming a quasi-flat band configuration. Zero energy corresponds to the InAs conduction band edge. The red solid line is the Fermi level $E_f$. $E_i$: $i^{th}$ electron subband, $HH_i$: $i^{th}$ heavy hole subband.

A small reduction of the Shubnikov-de-Haas frequency and thus of the n-type doping is observed with increasing Mn content. This is consistent with the calculated band structure shown in Fig. 3(c), if one simply assumes a reduction of the Fermi level position with increasing Mn content. $E_f$ drops from 290meV for the pure sample, close to 220meV for sample 630. For that sample, the charge density is found to be $n=3.0\times10^{12}cm^{-2}$. This reduction is also consistent with the established knowledge that Mn introduces acceptor levels into III-V semiconductors, [25] and with previously reported p-type conduction in InAs/GaSb at high Mn content. At dilute levels, Mn acceptor states will hence compensate the native n-doping of InAs.

**Cyclotron resonance at various Mn contents**

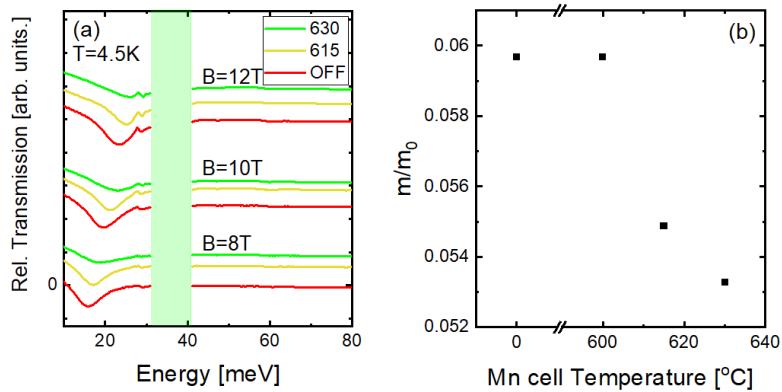

**FIG 4.** Cyclotron resonance measurements in the far-infrared at different magnetic fields, at 4.2K. The green box represents the reststrahlen band of the GaAs substrate. (b) Cyclotron masses extracted for the 4 samples from the cyclotron resonance at 12T.

To further corroborate this picture, cyclotron resonance measurements are carried out at 4.2K in the far-infrared. The optical transmission in the far-infrared for various fields measured for three samples (600, 615, and 630) is shown in Fig. 4(a). A minimum that shifts to higher energy versus magnetic field is observed indicating that the minimum is related to Landau levels. The position of the minimum is consistent with what is expected for the cyclotron resonance of the 2DEG for high Landau indices. The cyclotron mass extracted from this data at 12T is shown for various Mn contents in Fig. 4(b). The resonance yields an effective mass close to $0.059m_0$ for samples 0 and 600 consistent with what is expected for when $E_f$=290meV. This value is higher than what was previously reported, [16] [9] simply because the position of the Fermi energy is higher above the conduction band edges. At the same time, the energy of the transmission minimum increases with increasing Mn content indicating that the cyclotron mass is decreasing with increasing Mn concentration.

These measurements confirm that the quantum oscillations are due to the interfacial 2DEG of InAs/GaSb with the Fermi level crossing electron-subbands of comparable effective mass. As more Mn is introduced, the cyclotron mass decreases consistent with a slight lowering of the Fermi level deduced from Shubnikov-de-Haas measurements. One expects the effective mass to depend on energy in this system, since the bands are highly non-parabolic away from the band edges. From these measurements, it is thus evident that a 2DEG yields quantum oscillations in $In_{1-x}Mn_xAs$, in the presence of paramagnetism. Its charge density and the effective mass drop as more Mn is introduced into the system.

**Coexistence of electrons and holes in Mn-doped InAs/GaSb**

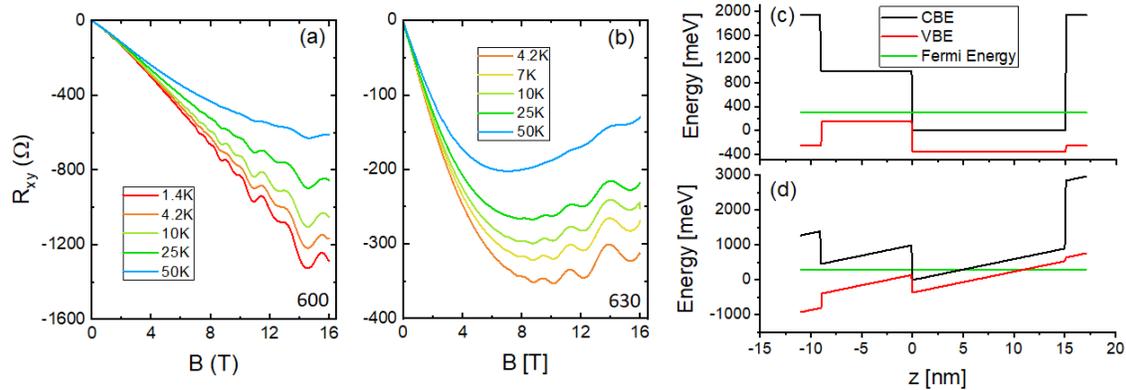

**FIG 5.** Hall effect versus increasing temperature in sample 600 (a) and 630 (b). (c) Band alignment of InAs/GaSb without Mn. The position of the Fermi level is fixed to be consistent with our quantum oscillations data. (d) Hypothetical sketch of the band alignment for Mn-doped InAs/GaSb. The Fermi energy has to cross the valence band away from the InAs/GASb interface, but the band offset has to be respected at the interface. CBE: conduction band edge, CBE: valence band edge.

The Hall effect measured up to 16T shown in Fig. 2 and 5 is inconsistent with that picture when Mn is introduced. Without Mn, InAs is lightly n-type but the $E_f$ is well above the valence band edge HH1 of GaSb as shown in Fig. 3(c) and $n_H$ agrees with the total $n$ extract from quantum oscillations. The Hall voltage becomes non-linear and its slope changes sign when the Mn content is increased as seen in Fig. 5(a,b) and Fig. 2. This indicates the presence of a finite density of holes in the system. In two samples, 600 and 630, quantum oscillations persist up to 50K as shown in Fig. 5(a,b). In sample 600, the Hall effect has a negative slope at 1.5K, but slowly develops a non-linearity with increasing temperature (Fig. 5(a)). In sample 630, the Hall effect is non-linear at 1.5K. Its slope is always negative at low field, but a positive slope develops at high field and gets enhanced with increasing temperature (Fig. 5(b)). The non-linearity of the Hall effect upon the introduction of Mn indicates the presence of a p-type conduction channel. We note here that the non-linearity of the Hall effect is unlikely to be caused by the anomalous Hall effect. This is because as temperature increases, the non-linearity is enhanced, contrary to what is expected to happen for an anomalous Hall effect resulting from magnetism.

At the Fermi energies extracted for the interfacial 2DEG, we do not expect $E_f$ to cross the valence subbands of GaSb in our model that assumes flat-band conditions (Fig. 5(c)). Let us consider the case of sample 630 with the highest Mn concentration. The quantum oscillations observed in that sample yield a Fermi wavevector consistent with what is obtained from the calculation (Fig. 3(c)) that assumed the known band-alignment of InAs/GaSb (Fig. 5(c)), for $E_f$=220meV. However, the observation of a hole gas in the Hall effect is not consistent with these results. $E_f$ remains too high to yield any crossing with the GaSb valence subbands at 220meV. This fact leads us to conclude that the flat band assumption used for undoped InAs/GaSb no longer holds when Mn is introduced. At high Mn content, InAs likely becomes p-type far away from the interface as Mn yields impurity levels near its valence band edge. [20] [26] The Fermi level should thus lie near the valence band edge of InAs away from its interface with GaSb. A substantial band bending effect could result from this, to preserve the band offset at the interface with GaSb as shown in Fig. 5(d). This hypothetical alignment can yield a situation that maintains the quantum oscillations resulting from the 2DEG while introducing a low mobility hole gas away from the interface, that only impacts diffusive transport such as the normal Hall effect.

### III. Conclusion

From this study, we conclude that Mn indeed impacts the band alignment and electronic structure of InAs/GaSb, beyond simply introducing magnetic exchange and reducing the charge density. At low Mn content, $In_{1-x}Mn_xAs$/GaSb samples host a coexisting 2DEG and hole gas that is due to band bending caused by the Mn impurity levels in InAs. We note that paramagnetism was measured in sample 630, but we could not detect the effect of magnetic exchange splitting in Shubnikov-de-Haas measurements, since the Fermi level crosses high index Landau levels at 16T (n>4). But importantly, we have shown that Mn allows the tuning of the Fermi level and alters the band alignment in InAs/GaSb. Its action is highly non-trivial and fundamentally interesting. Its understanding requires a better knowledge of the charge distribution and the band bending profile of the system to enable reliable self-consistent calculations. The realization of a QAHE in this structure can only be achieved if the action of Mn is properly quantified and controlled, since this band bending can generate additional diffusive conduction channels that will short out the chiral edge states.

**Acknowledgements.** Work support by NSF-DMR-1905277. We also acknowledge support from ANR grant N°-19-CE30-022-01. We thank Mansour Shayegan for useful conservations.

(2008).

[26]    M. K. Parry and A. Krier, J. Cryst. Growth **139**, 238 (1994).